# Load Balancing in a Networked Environment through Homogenization

Mahmud Shahriar HOSSAIN
Department of Computer Science and Engineering, Shahjalal University of Science and Technology
Sylhet-3114, Bangladesh. E-mail: shahriar-cse@sust.edu

M. Muztaba FUAD
Department of Computer Science, Montana State University
Bozeman, MT 59717, USA. E-mail: fuad@cs.montana.edu

Debzani DEB
Department of Computer Science, Montana State University
Bozeman, MT 59717, USA. E-mail: debzani@cs.montana.edu

Kazi Muhammad Najmul Hasan KHAN
Department of Computer Science and Engineering, Shahjalal University of Science and Technology
Sylhet-3114, Bangladesh. E-mail: najmul_bd@yahoo.com

and

Dr. Md. Mahbubul Alam JOARDER
Institute Of Information Technology (IIT), University of Dhaka
Dhaka-1000, Bangladesh. E-mail: joarder@udhaka.net

## ABSTRACT

Distributed processing across a networked environment suffers from unpredictable behavior of speedup due to heterogeneous nature of the hardware and software in the remote machines. It is challenging to get a better performance from a distributed system by distributing task in an intelligent manner such that the heterogeneous nature of the system do not have any effect on the speedup ratio. This paper introduces homogenization, a technique that distributes and balances the workload in such a manner that the user gets the highest speedup possible from a distributed environment. Along with providing better performance, homogenization is totally transparent to the user and requires no interaction with the system.

## 1. INTRODUCTION AND BACKGROUND

Distributed computing in a LAN environment encompasses heterogeneous infrastructure due to the variety of hosts in processing power, hardware architecture, memory, resident operating system, background daemons etc. Simply allotting equal amount of workload to each machine would cause speedup degradation with the appearance of low-performance machines. Although many systems have been proposed for distributed computing, very few operates efficiently in this type of heterogeneous environment. Moreover as speedup is always dependent on actual computation time and the overhead for communication, a profound relationship between load, computation and data to be transferred is mostly essential. By establishing this relationship homogenization enriches communication abstraction.

System such as AdJava [3, 4] harnesses the computing power of underutilized hosts across a LAN or WAN, provides load balancing but suffers from penalty of migration time of the object and is useful only for scientific applications. Nieuwpoort et al. [5] established a divide-and-conquer model for writing distributed supercomputing applications on hierarchical wide-area systems. But the divide-and-conquer strategy may result in high round-trip time. Although there are several distributed systems [1, 2, 7, 9], there is hardly any work on intelligent job distribution and load balancing.

Triangular Dynamic Architecture (TDA) [6] introduces a mechanism of distributed processing and parallel computation for balancing the workload among the idle machines of a network. An intelligent server must divide the requested jobs efficiently so that the distribution mechanism properly balances the load across the system. TDA provides a dynamic nature of distributed and parallel processing that possesses platform independence and a load-balancing tool called homogenization. Homogenization is a process that assures TDA to balance the workload across a networked environment in a dynamic and intelligent way. Homogenization considers the heterogeneous parameters of a LAN and allocates proportional amount of workload to the clients. Thus, Homogenization assures speedup even when low-performance machines are involved.

Homogenization can be applied to any kind of linearly divisible distributed processing system to ensure better performance. It can be used for solving computation intensive scientific problems. Commercial applications, where millions of records are manipulated and thousands of clients are to be served, homogenization would provide better performance.

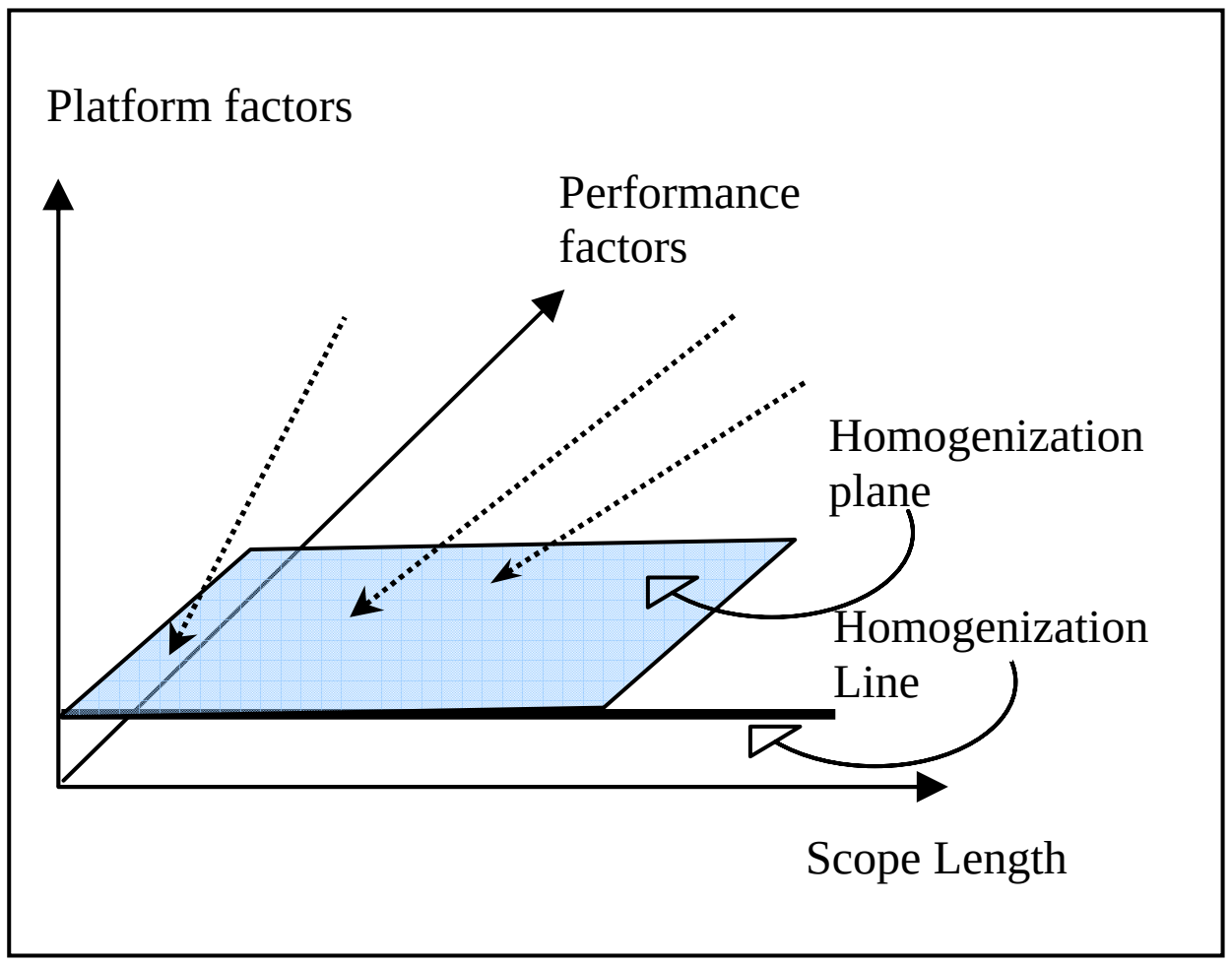

**Figure 1: Illustration of homogenization techniques for TDA.**

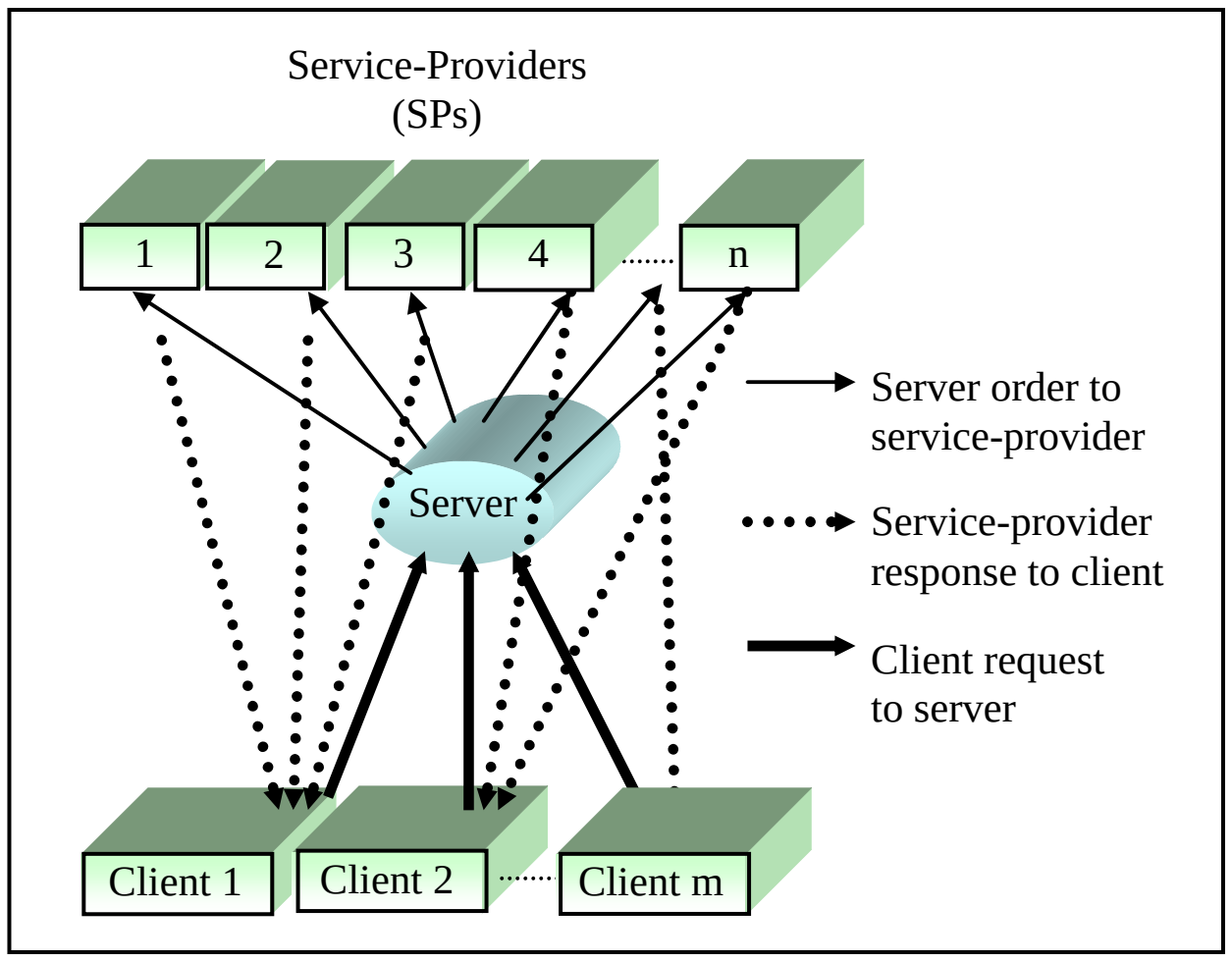

**Figure 2: Triangular Dynamic Architecture.**

## 2. SYSTEM ARCHITECTURE

Homogenization is a mechanism that is established over Triangular Dynamic Architecture (TDA). TDA server, service provider and client. Homogenization facilitates TDA in distribution of workload in a load-balancing manner.

**Homogenization**

Figure 1 illustrates the homogenization process for TDA. Java Virtual Machine (JVM) [8] brings all the hosts in TDA to the same platform named homogenization plane. This is the first level of homogenization. In the homogenization plane all the machines are of same virtual platform but they are of different performance factors. The TDA server performs the next level of homogenization. It brings the service-providers to the homogenization line. This level of homogenization is performed by allotment-variation of workload depending on the performance factors of the service-providers.

In the homogenization line, all the service-providers take same amount of time to complete corresponding sub-requests. Scope length is the length of allotment of workload to a service-provider decided by the server. Scope length variation makes all the service-providers finish their computation at the same time.

Homogenization promotes the way of establishing triangular relationships in TDA. The client sends request to the TDA server, the server thereafter granulizes the request into subparts and sends them to service-providers. If the requests were divided into equal pieces, slower service providers would take more time than a faster one. Homogenization allows granulation of load in a balancing manner so that all the service providers take same amount of time to compute their corresponding subparts. That is homogenization improves the mechanism of distribution from the server in an intelligent way, which enables load-balancing technique. A slower machine now gets smaller portion of the computation compared to a faster one. The distribution depends on a dynamic collection of performance from the service providers.

The server maintains several tables in its local database that helps distributing the load. The server actually calculates the scope-length to be offered to a particular service-provider, using the tables of the local database. Most critical knowledge-issues are performance of the service-providers, their response time, list of services provided by a service-provider, etc. A background process in the service-provider informs the server about its current load after certain time interval. The server maintains this information and based on the stored information, the server generates a performance number, which is called the homogenized performance. Homogenized performance is the outcome of the second level homogenization of Figure 1. That is, the first level of homogenization, bringing all the machines to homogenization plane is done by JVM, where TDA server with the generation of homogenized performance value for each machine performs the second level of homogenization, bringing all the service-providers to homogenization line. Thereafter, the server depends on the homogenized performance of the service-providers for the balanced distribution of load.

Figure 2 illustrates a sample of TDA that shows multiple triangular relationships established against a request from the client. The relationships are established dynamically based on the homogenized performance of the service-providers.

**Mathematical synopsis**

Let the time to complete a job in a standalone machine be $T$; if there exists $N$ computers those have the same performance along with same homogeneous behavior as the standalone one, then total time to compute the same job with ***N*** computers will be

$$T_N = \frac{T}{N} \cdots\cdots\cdots\cdots\cdots\cdots\cdots\cdots\cdots\cdots\cdots (1)$$

In fact, this is the best theoretical case where the distribution mechanism itself does not carry any cost of distribution. If the cost of distribution is ***O(L)*** where ***L*** is the amount of load that is to be distributed among ***N*** computers, total time to compute the same job would be

$$T_N = \frac{T}{N} + O(L) \cdots\cdots\cdots\cdots\cdots\cdots\cdots\cdots (2)$$

So far, the mathematical synopsis is based on a homogeneous environment. But a real networked environment is composed of heterogeneous elements. TDA server depends on the performance values of the service-providers for a balanced distribution. As a result the concept of "number of computers" becomes somewhat different. The number of computers depends on two parameters. They are: $P_S$, the performance of the standalone machine and $P_T$ , the sum of performance values of all the machines invoked by the distribution. That is,

$$P_T = \sum_{i=1}^{N} P_i \cdots\cdots\cdots\cdots\cdots\cdots\cdots\cdots\cdots\cdots (3)$$

Hence, although physically the number of computers is ***N***, during homogenization virtually the number of computers would become,

$$N_H = \frac{P_T}{P_S} = \frac{\sum_{i=1}^{N} P_i}{P_S} \cdots\cdots\cdots\cdots\cdots\cdots\cdots\cdots (4)$$

Thereafter for homogenization in a heterogeneous infrastructure Eq. (2) would get the following form:

$$T_{NH} = \frac{T}{N_H} + O(L) = \frac{T}{\frac{\sum_{i=1}^{N} P_i}{P_S}} + O(L) \cdots\cdots\cdots (5)$$

Naturally, speedup for a homogenized system can be defined as

$$S_{NH} = \frac{T}{T_{NH}} = \frac{T}{\frac{T}{\frac{\sum_{i=1}^{N} P_i}{P_S}} + O(L)} \cdots\cdots\cdots\cdots\cdots (6)$$

The distribution overhead ***O(L)*** would become negligible for high degree of computation where actual computation time, $\frac{T}{N_H}$ plays the most dominating role. In this case,

$$T_{NH} = \frac{T}{N_H} \cdots\cdots\cdots\cdots\cdots\cdots\cdots\cdots\cdots\cdots\cdots (7)$$

and

$$S_{NH} = \frac{\sum_{i=1}^{N} P_i}{P_S} \cdots\cdots\cdots\cdots\cdots\cdots\cdots\cdots\cdots\cdots (8)$$

This case would lead to a linear increment of speedup in accordance with $N_H$, the virtual number of computers.

Overhead ***O(L)*** is not only a function of communication time but also it depends on the time to deliver the information about the network. As the information is stored in the server, the manipulation of the distribution objects takes a very low amount of time compared to the communication time. As a result overhead is a very linear function of only the load i.e., ***O(L)***.

$$O(L) = M \times L \cdots\cdots\cdots\cdots\cdots\cdots\cdots\cdots\cdots\cdots (9)$$

***M***, the slope of the line depends on the infrastructure of the local are network, to which the system is established.

## 3. PERFORMANCE ANALYSIS

To verify the potentiality of homogenization, a scientific application is implemented in TDA. Performance is measured in two types of environment: heterogeneous environment, and homogenized environment. A homogenized environment is one where TDA has applied homogenization i.e. in reality homogenized environment is a heterogeneous one, but TDA homogenized the overall system.

Matrix multiplication is a common scientific computation that is to be solved for different scientific problems. Considering the simplest algorithm that multiplies two matrices with three loops, the experiment is performed. All the statistics taken are for the same network, same service-providers and the same thin client, as well as the same TDA server. For experimental purpose, the test matrices were all square matrices. Every time two square matrices of same size were requested to the server to distribute.

Only the first matrix is granulized into pieces and sent to different service-providers. Each service-provider gets a copy of the second matrix from the thin client. Each service-provider then calculates a portion of the result and sends it directly to the thin client that requested for the job. The thin client combines the result when all the portions are received.

The experiments are taken with various combinations of Intel machines. They were varying in CPU speed, memory size, operating system, user processes, background daemons and many other parameters. Pentium II, III, and IV Intel machines with physical memory ranging from 64 to 128 MB are used. All of them are connected by 100 Mbps Ethernet network. All the TDA components were running over the Virtual Machine provided by SUN's JDK version 1.2.2 or higher.

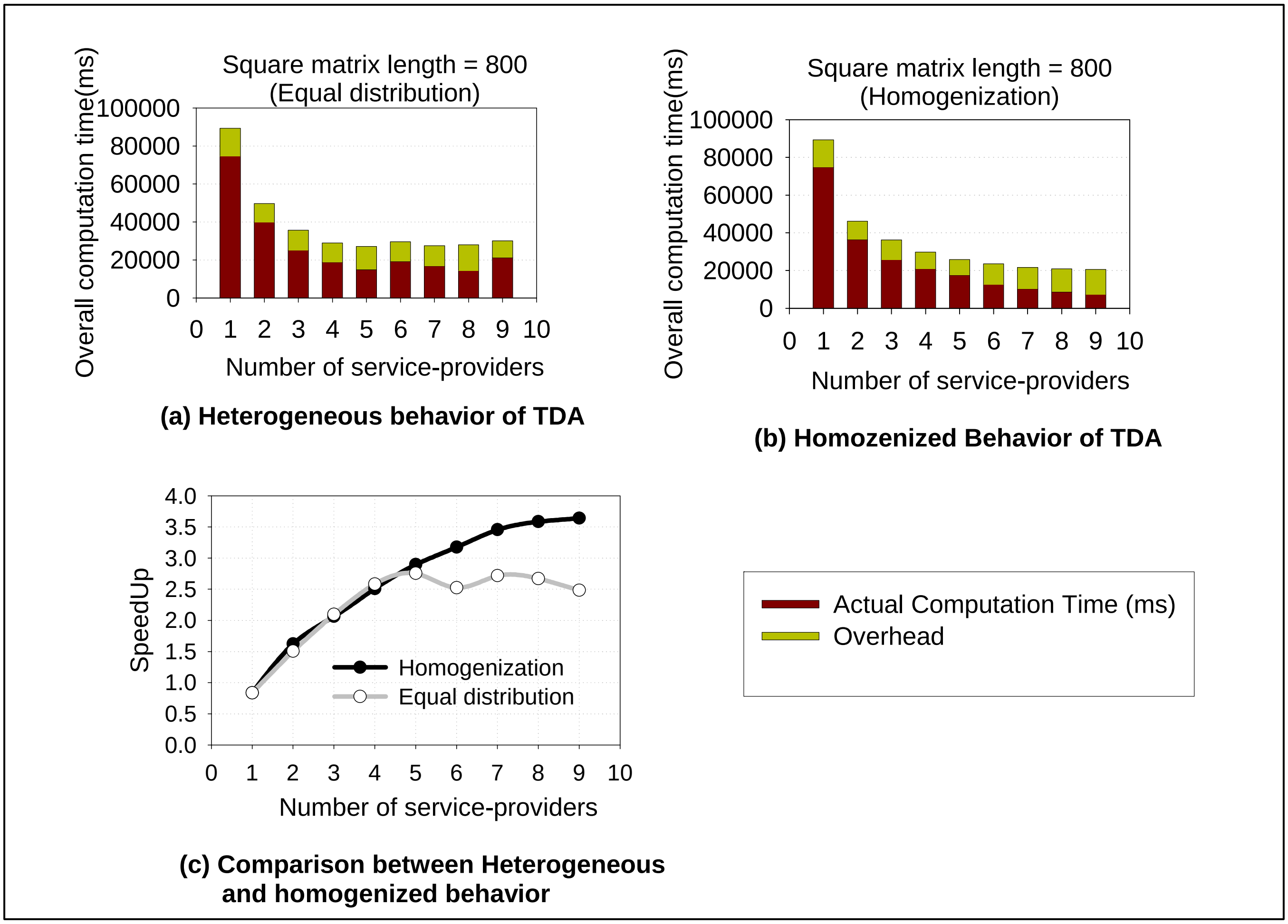

**(a) Heterogeneous behavior of TDA**

**(b) Homozenized Behavior of TDA**

**(c) Comparison between Heterogeneous and homogenized behavior**

**Figure 3: Performance Analysis for homogeneous and homogenized behavior.**

**Heterogeneous behavior of TDA**

Figure 3 shows both heterogeneous and homogenized behavior of TDA for square matrix size of 800. The dark portion of a bar indicates the actual computation time and the grey part represents the overhead due to communication distance. From Figure 3(a), it is evident that introduction of successive service-providers reduces the actual computation time. Closer inspection shows that introduction of the sixth and the ninth service-providers does not reduce the actual computation time, rather computation time is increased. This type of degradation of performance is found because the sixth and ninth service-providers were comparatively of low CPU speed. Equal allotment of load results in heterogeneous pattern of speedup. The heterogeneous pattern of speedup is shown in Figure 3(c) with a gray line. The speedup pattern shows that speedup is decreased when sixth and ninth service-providers are involved. Overhead affects speedup because overall computation time is composed of actual computation time and overhead. Overhead is an additive function of communication time and decision making time of the server.

**Homogenized behavior of TDA**

The same analysis is taken with the only exception that now allotment of load is not equal. TDA homogenized the environment. The physical environment is the same as heterogeneous one, but now homogenization is applied. Figure 3(b) shows that application of homogenization assures decrease in actual computation time although the infrastructure is heterogeneous. Corresponding speedup is shown in Figure 3(c) with a black line.

Figure 4 compares the same homogenized behavior of Figure 3 with formulae of section 2. The summation of performances of the machines is taken from a single snapshot. But runtime performance varies during operation. Moreover the overhead may vary time to time due to congestion in the network. As a result, a deviation from the formula is found for the speedup. The overhead function is a linear one. $\boldsymbol{M}$, of Eq. (9) is 20 for our network. It can vary network to network depending on the speed of the network.

Introduction of newer service-providers causes speed-up improvement regardless their configuration. But the acceleration of speedup is decreased while large amount of service-providers is involved in a distribution. This clarifies that the almost constant overhead becomes pronounced when the actual computation time is reduced. Subsequent involvement of too many service-providers results in slow speedup improvement. In this experiment, Figure 3(c) shows that homogenization provides a maximum speedup of 3.6 with nine service-providers but non-homogenized distribution provides maximum speedup of 2.8 with 5 service-providers.

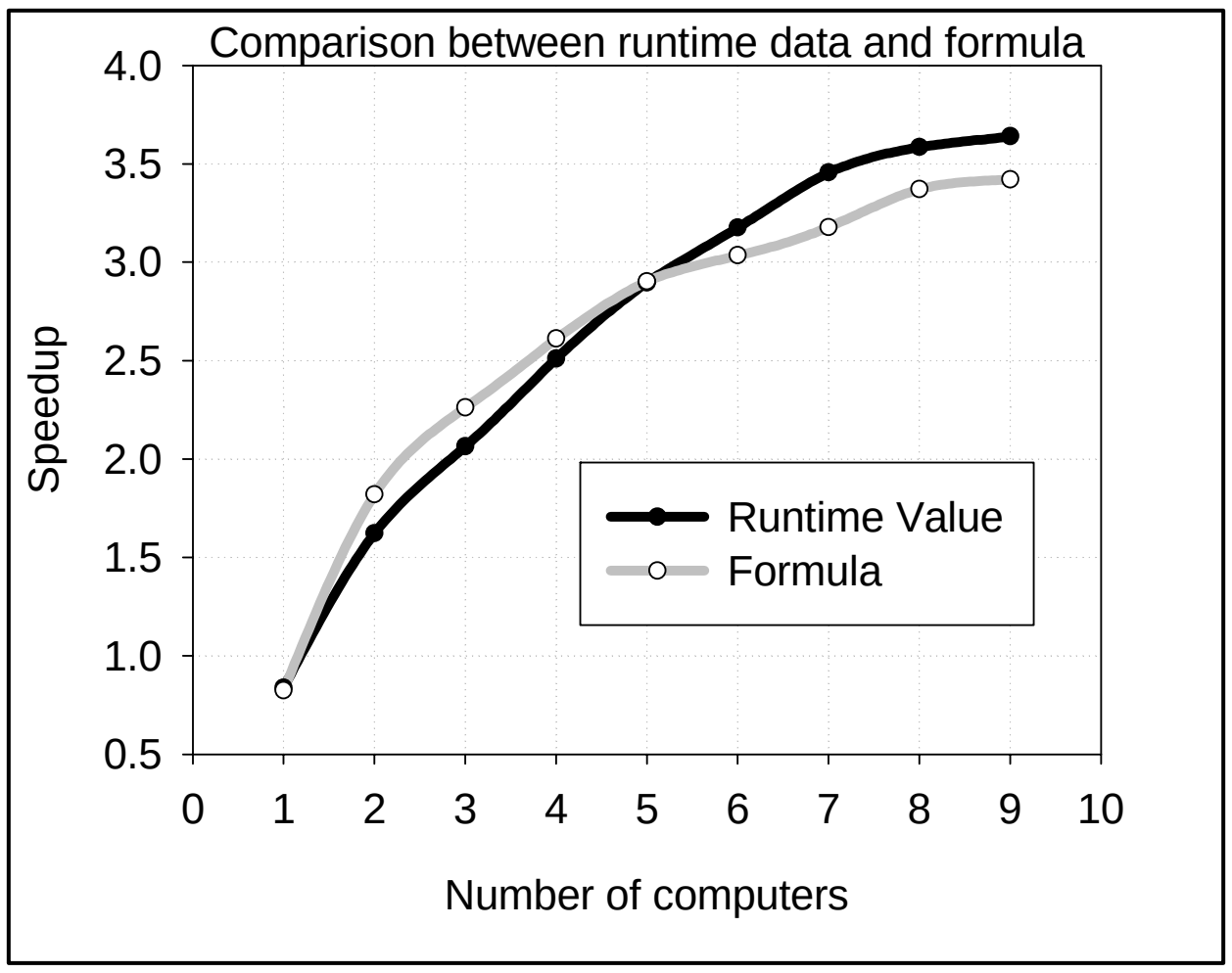

**Figure 4: Performance analysis for homogenized behavior of TDA and performance by formula.**

**Load and linearity**

Speedup also depends on the size of the load. Different sizes of matrices are used to understand the behavior. Figure 6(a) shows the speedup lines at different size of matrix multiplication. The figure depicts heterogeneous performance improvement. The matrix sizes are 200, 400, 600, 800 and 1000. For some of the sizes, speedup is less than unity, which illustrates that TDA could not improve the performance because the load was too small. In this case, overheads dominate over the actual computation time. During the size 200, such degradation is found. For all other sizes, speedup is greater than unity. It proves

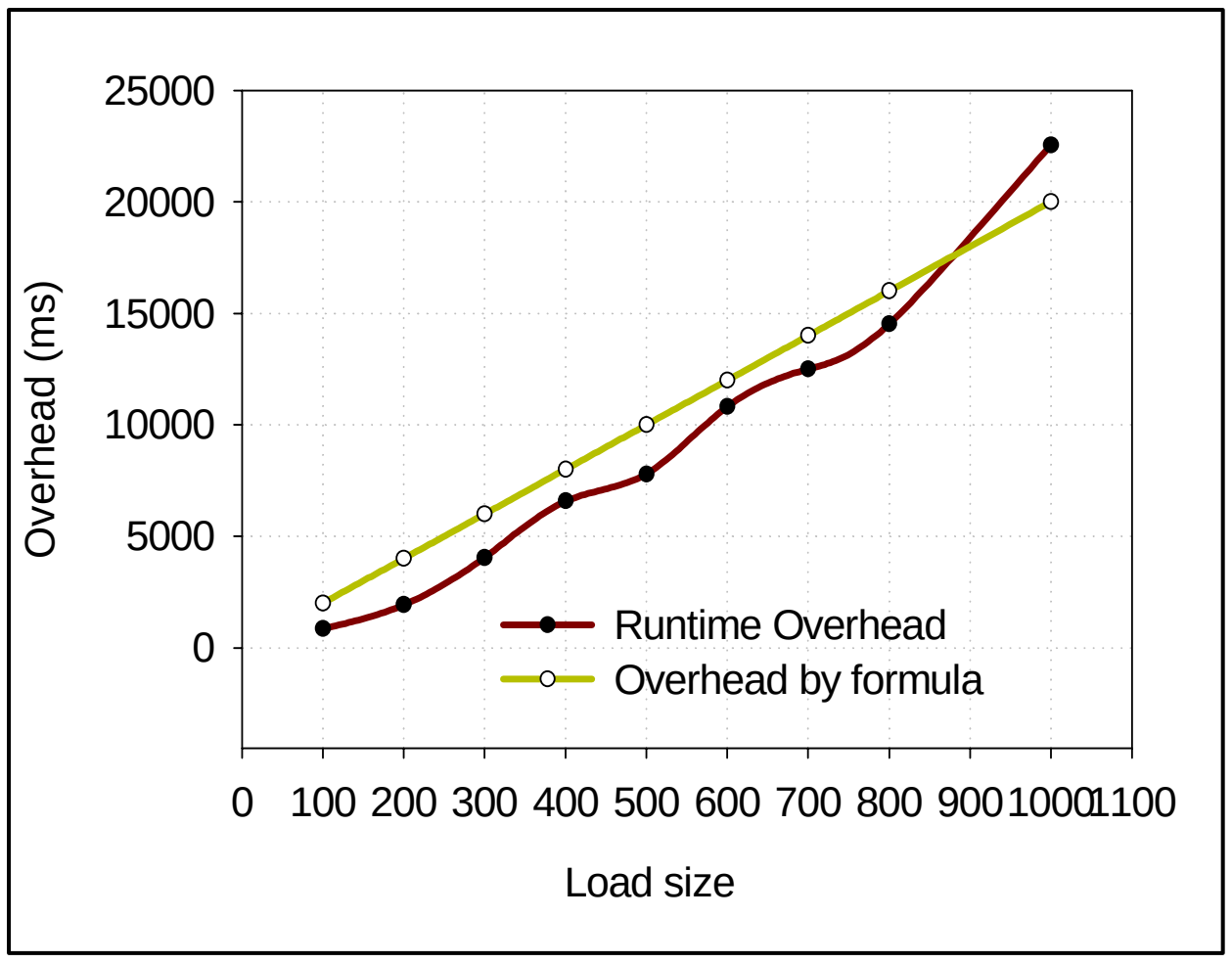

**Figure 5: Overhead pattern for different load.**

that TDA shows higher performance at higher degree of load.

Figure 6(b) depicts that when the load size is 1000, the speedup line is almost linear, while at lower sizes speedup is more nonlinear.

This proves that overhead becomes negligible for huge amount of load hence speedup becomes a linear function. The corresponding homogenized performance for the same heterogeneous infrastructure is given in Figure 6 (b), which shows steady improvement of performance at higher amount of loads. A comparison between Figure 6 (a) and Figure 6 (b) shows that the maximum speedup reached during non-homogenized situation is around 3.5 where the maximum speedup reached during homogenization is around 5.5 which describe the nobility of homogenization through TDA.

The overhead function is a linear one. ***M***, of Eq. (9) is 20 for our network. Formula overhead and real overhead for different load size are sketched in Figure 5.

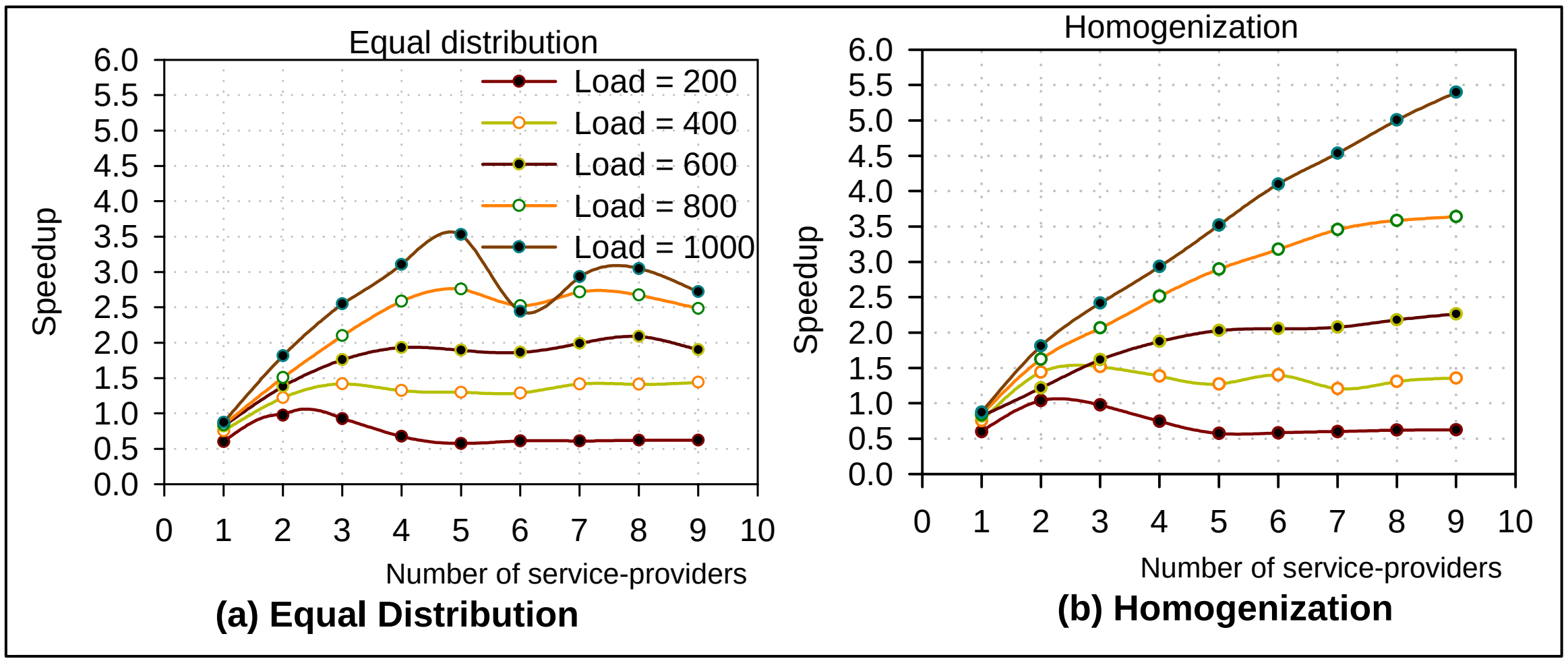

**Figure 6: Speedup with different loads.**

## 4. CONCLUSION

Homogenization is applicable to linearly divisible problems. It can be used for solving computation intensive scientific problems. Commercial applications, where millions of records are manipulated, thousands of clients are to be served, homogenization would provide better performance. Homogenization does not require any kind of user interaction for its knowledge centric distribution mechanism. When someone is exploring the space, it becomes challenging to detect which part of the space is poor and which part is rich. Homogenization enables system independent performance evaluation, which is preserved by the TDA server. As a result homogenization would become applicable in wide range of problem space. Once again, if the work to be distributed is not a linearly divisible one, homogenization can be used to enrich the system with Migration of Thread [10] from the weaker part to the rich part of the problem space. The future direction of the work is to provide Distributed Artificial Intelligence (DAI) with the environment of Distributed Intelligent System (DIS) to secure maximum possible speedup not only homogenized by the server but also gained by intelligent service providing agents.

Homogenization technique based of TDA, is an enriching mechanism of job distribution. TDA granulizes computation intensive jobs to concurrent pieces using homogenization and operates them in a dynamic environment to reduce total processing time. Experimental analysis shows that in a heterogeneous environment, homogenization provided a 55% increase in speedup relative to maximum non-homogenized performance. It is established with an automatic manner in TDA as a transparent load-balancing tool.

Homogenization provides better processing time in a distributed computing environment. For implementing homogenization, the present JVM remains unchanged. The current implementation is fully based on the existing JVM and that way TDA fulfills its main goal of providing a distributed computing environment in an existing LAN.